\documentclass[aps]{revtex4}
\usepackage{eurosym}
\usepackage{amsfonts}
\usepackage{amsmath}
\usepackage{amssymb,epsf}
\usepackage{color}
\usepackage{graphicx}
\usepackage{epstopdf}
\usepackage{float}
\usepackage{caption}
\usepackage{subfig}

\begin{document}

\title{The structure of hybrid neutron star in Einstein-$\Lambda$ gravity}
\author{T. Yazdizadeh$^{1}$, G. H. Bordbar$^{2,3}$\footnote{
email address: ghbordbar@shirazu.ac.ir}, and B. Eslam Panah $^{4,5,6}$\footnote{
email address: eslampanah@umz.ac.ir}}
\affiliation{$^{1}$Islamic Azad University, Bafgh Branch 89751-43398, Bafgh, Iran\\
$^{2}$ Department of Physics, Shiraz University, Shiraz 71454, Iran\\
$^{3}$ Department of Physics and Astronomy, University of Waterloo, 200
University Avenue West, Waterloo, Ontario N2L3G1, Canada\\
$^{4}$ Department of Theoretical Physics, Faculty of Science, University of
Mazandaran, P. O. Box 47416-95447, Babolsar, Iran\\
$^{5}$ICRANet-Mazandaran, University of Mazandaran, P. O. Box 47416-95447,
Babolsar, Iran\\
$^{6}$ ICRANet, Piazza della Repubblica 10, I-65122 Pescara, Italy.}

\begin{abstract}
In this paper, we investigate the structure of neutron stars by considering
both the effects of the cosmological constant and the existence of quark
matter for neutron stars in Einstein's gravity. For this purpose, we use a
suitable equation of state (EoS) which includes a layer of hadronic matter,
a mixed phase of quarks and hadrons, and a quark matter in the core. To
investigate the effect of the cosmological constant on the structure of
hybrid neutron stars, we utilize the modified TOV equation in Einstein-$%
\Lambda $ gravity. Then we drive the mass-radius relation for different
values of the cosmological constant. Our results show that for small values
of the cosmological constant ($\Lambda $), especially for the cosmological
constant from the cosmological perspective $(\Lambda =10^{-52}$ $m^{-2})$, $%
\Lambda $ has no significant effect on the structure of hybrid neutron
stars. But for higher values, for example, by considering $\Lambda >10^{-14}$
$m^{-2}$, this quantity affects the maximum mass and radius of these stars.
We find an upper limit for the cosmological constant as $\Lambda <9\times
10^{-13}m^{-2}$, based on the fact that the gravitational redshift cannot be
more than $1$ for stars. The maximum mass and radius of these stars decrease
by increasing the cosmological constant $\Lambda$. Also, by determining and
analyzing radius, the compactness, Kretschmann scalar, and gravitational red
shift of the hybrid neutron stars with $M=1.4M_{\,\odot }$ in the presence
of the cosmological constant, we find that by increasing $\Lambda $, they
are contracted. Also, our results for dynamical stability show that these
stars satisfy this condition.

\noindent \textbf{Keywords}: Hybrid neutron star, Structure, Einstein-$%
\Lambda $ gravity, Cosmological constant
\end{abstract}

\maketitle

\section{Introduction}

Einstein gravity is a successful theory of gravity, especially for
explaining the motion of planets and stars at the macroscopic scale, the
solar system phenomena, and light bending to precision measurements of the
orbits of binary pulsars \cite{Kramer}, so far since it is strongly valid
under the weak gravitational field approximation. In addition, the most
recent cosmological observations are consistent with the standard
cosmological models built on Einstein's gravity. Another powerful
prediction of Einstein gravity is related to the presence of gravitational
waves, which was detected by the LIGO detections in 2016 \cite{Abbott}.
However, there is a mysterious late-time acceleration phase in our Universe.
Indeed, the observations of Supernova type Ia (SNeIa) showed that the
expansion of our Universe is currently undergoing a period of acceleration 
\cite{PerlmutterI,Riess}. The Einstein gravity cannot describe this
acceleration. In order to explain this acceleration there are some modified
theories of gravity (see refs. \cite%
{modI,modII,modIII,modV,modVI,modVII,modVIII}, for more details). It
is notable that the realization of gravitational wave astronomy provides us
with the possibility of discriminating among Einstein gravity and other
gravity theories \cite{Corda2009}. In other words, by improving the results
of the LIGO detections, we can determine the validation of Einstein gravity
extensions. Also, the future of gravitational theories in the framework of
gravitational wave astronomy after the recent GW detections is discussed in
ref. \cite{Corda2018}. Among these modified gravities, adding a
(cosmological) constant ($\Lambda $) to Lagrangian of Einstein gravity is a
simplified theory to explain this acceleration \cite{Padmanabhan,Frieman}
(Historically, Einstein presented this constant to save the universe from
expanding, but then he rejected it after discoveries by Hubble).

Because of the high density in the core of neutron stars, which is one of
the densest objects in the universe, a quark matter is expected in their
interior \cite{Ivanenko,Itoh,Fritzsch,Collins}. This system with such high
density is the interest and favorite case for physicists and astrophysicists 
\cite%
{NSQM1,NSQM2,NSQM3,NSQM5,NSQM7,NSQM8,NSQM9,NSQM10,NSQM11,NSQM13,NSQM14,NSQM15,NSQM16,NSQM17,NSQM18,NSQM20,NSQM22,NSQM23,NSQM24}%
. There are a lot of uncertainties in the calculations and compositions of
the neutron star. Thus people use different models and various equations of
state to compare their results with the observational data of neutron stars.
In this work, we intend to obtain the structure properties of a neutron star
with a quark core which is named hybrid star. For this purpose, we assume a
hybrid star composed of three parts: a layer of hadrons in the surface, a
hadron-quark mixed matter in the middle, and a pure quark matter in the
center of the star.

To determine the properties of stars, we need a hydrostatic equilibrium
equation (HEE) that satisfies static gravitational equilibrium. The TOV
equation was the first HEE equation used to calculate the structure of
stars. It was derived in Einstein gravity by Tolman, Volkoff and Oppenhimer 
\cite{Tolman1934, Tolman1939, Oppenheimer}. Many authors have been studied
compact stars by using TOV equation \cite%
{TOV1,TOV2,TOV3,TOV4,TOV5,TOV7,TOV8,TOV9}. It is worth mentioning that, for
studying compact objects such as neutron stars in modified theories of
gravity, we must extract the modified TOV equation. For example, the
modified TOV equation in dilaton gravity \cite{Dilaton},
vector-tensor-Horndeski theory of gravity \cite{Momeni}, gravity's rainbow 
\cite{RainbowI,RainbowII}, $F(R)$ and $F(G)$ gravities \cite%
{F(R)I,F(R)II,F(R)III}, massive gravity \cite{MassiveI,MassiveII}, mimetic
gravity \cite{mimeticI,mimeticII}, and $F(R,T)$ gravity \cite{Rahaman}, have
been evaluated (see \cite%
{NS1,NS2,NS4,NS5,NS6,NS7,NS8,NS10,NS11,NS12,NS13,NS14,NS15,NS16}, for more
details). Therefore, we have to modify the TOV equation to study the
structure of neutron stars with a quark core in Einstein-$\Lambda $ gravity.

In this paper, we use a suitable EoS which includes three parts, a hadronic
matter layer (LOCV Method \cite{bordbar1998}), a mixed part of quarks and
hadrons (with Gibss conditions), and a quark matter in core (MIT bag model)
where studied in some literature {\cite{EoSI,EoSII}}. The mathematical form
of this EoS presented as a polynomial function, 
\begin{equation}
P=\sum_{i=1}^{7}a_{i}\mathcal{E}^{7-i},  \label{Prho}
\end{equation}%
in which $a_{i}$ are as 
\begin{eqnarray}
a_{1} &=&1.194\times 10^{-57},~\ \ \&~~~a_{2}=-0.246\times 10^{-40},  \notag
\\
a_{3} &=&2.011\times 10^{-25},~~~\&~~~a_{4}=-8.123\times 10^{-10},  \notag \\
a_{5} &=&1.656\times 10^{6},~~~\ \ \ \&~~~a_{6}=-1.201\times 10^{21},  \notag
\\
a_{7} &=&2.915\times 10^{35}.
\end{eqnarray}

It is worthwhile that energy conditions (the null energy condition, weak
energy condition, strong energy condition, and dominant energy condition at
the center of hybrid neutron star), stability, and Le Chatelier's principle
of this EoS have been investigated in ref. \cite{EslamYB}.

\section{Modified TOV equation in Einstein-$\Lambda $ gravity}

The action of the Einstein gravity with the cosmological constant in $4$%
-dimensions is given by 
\begin{equation}
I_{G}=\frac{1}{2\kappa }\int d^{4}x\sqrt{-g}(R-2\Lambda )+I_{Matt},
\label{eq16}
\end{equation}%
where $R$ is the Ricci scalar and $I_{Matt}$ is the action of matter field. $%
\kappa =\frac{8\pi G}{c^{4}}$, $c$ is the velocity of light and $G$ is the
Newtonian gravitational constant. Varying the action (\ref{eq16}) with
respect to the metric tensor $g_{\mu }^{\nu }$, the equation of motion can
be written as 
\begin{equation}
R_{\mu }^{\nu }+\frac{1}{2}Rg_{\mu }^{\nu }+\Lambda g_{\mu }^{\nu }=\frac{%
8\pi G}{c^{4}}T_{\mu }^{\nu },  \label{eq17}
\end{equation}%
where $R_{\mu }^{\nu }$ is the symmetric Ricci tensor. $T_{\mu }^{\nu }$ is
the energy-momentum tensor of perfect fluid as $T^{\mu \nu }=(c^{2}\epsilon
+P)U^{\mu }U^{\nu }-Pg^{\mu \nu }$ (where $\epsilon $ and $P$ are density
and pressure of the fluid which are measured by local observer,
respectively, and $U^{\mu }$ is the fluid $4$-velocity). Considering the
field equation (\ref{eq17}) and the mentioned energy-momentum tensor with a
spherical symmetric spacetime in the following form%
\begin{equation}
ds^{2}=c^2f(r)dt^{2}-\frac{dr^{6}}{g(r)}-r^{2}\left( d\theta ^{2}+\sin
^{2}\theta d\varphi ^{2}\right) ,  \label{eq18}
\end{equation}%
the HEE in Einstein-$\Lambda $ gravity is given \cite{BordbarHE}%
\begin{equation}
\frac{dP}{dr}=\frac{3c^{2}GM+r^{3}(\Lambda c^{4}+12\pi GP)}{%
c^{2}r[6GM-c^{2}r(\Lambda r^{2}+3)]}(c^{2}\epsilon +P),  \label{eq19}
\end{equation}%
where for $\Lambda =0$ this equation reduces to TOV equation (see \cite%
{BordbarHE}, for more details).

\section{Structure of Hybrid Neutron Stars in the Presence of the
Cosmological Constant}

\begin{figure*}[tbh]
\centering
\includegraphics[width=0.6\linewidth]{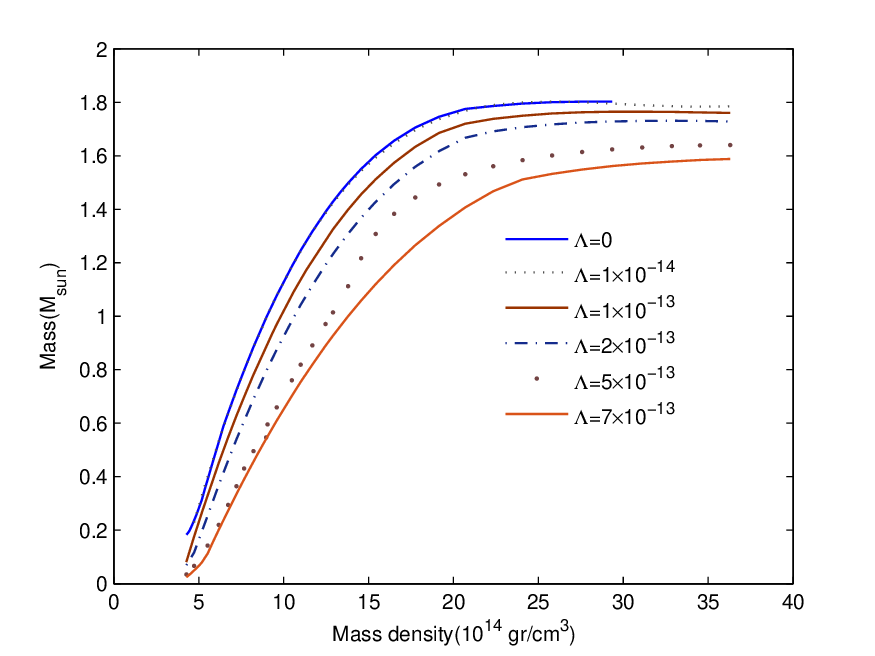}
\caption{Maximum gravitational mass for the hybrid neutron stars at
different $\Lambda$.}
\label{mass}
\end{figure*}

\begin{figure*}[tbh]
\centering
\includegraphics[width=0.6\linewidth]{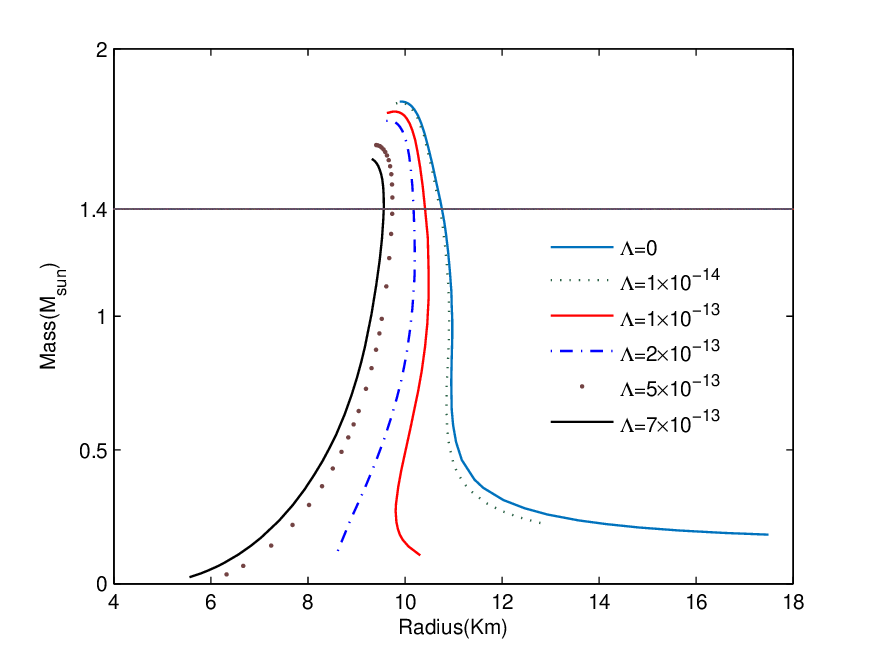}
\caption{Mass-radius diagram for the hybrid neutron stars at different $%
\Lambda$.}
\label{radius}
\end{figure*}

To determine structures of hybrid neutron stars in the presence of the
cosmological constant, we use the HEE in the Einstein-$\Lambda $ gravity and
the mentioned EoS in Eq. (\ref{Prho}). By numerical integration of this
equation, we determine the structure of hybrid stars. The results of maximum
gravitational mass and the radius-mass diagram for the hybrid neutron stars
are presented in Figs. \ref{mass} and \ref{radius}. %
%%%%%%%%%%%%%%%%%%%%%%%%%%%%%%%%%%%%%%%%%%%%%%%%%%%%%%%%%%%%%%%
%%%%%%%%%%%%%%%%%%%%%%%%%%%%%%%%%%%%%%%%%%%%%%%%%%%%%%%%%%%%%%%
%%%%%%%%%%%%%%%%%%%%%%%%%%%%%%%%%%%%%%%%%%%%%%%%%%%%%%%%%%%%%%%
%%%%%%%%%%%%%%%%%%%%%%%%%%%%%%%%%%%%%%%%%%%%%%%%%%%%%%%%%%%%%%%

Our obtained results for the maximum gravitational mass and the
corresponding radius, and also other structure properties of the hybrid
neutron stars have been presented in Table \ref{tab1}. %
%%%%%%%%%%%%%%%%%%%%%%%%%%%%%%%%%%%%%%%%%%%%%%%%%%%%%%%%%%%%%%%%%
\begin{table}[tbp]
\caption{Structure properties of hybrid neutron stars for different $\Lambda$%
.}
\label{tab1}
\begin{center}
\begin{tabular}{||c|c|c|c|c|c||}
\hline\hline
$\Lambda (m^{-2}) $ & $M_{max}\left(M_{\,\odot}\right) $ & $R\left(km\right) 
$ & $R_{Sch}\left(km\right)$ & $z$ & $M_{BB}\left(M_{\,\odot}\right) $ \\ 
\hline\hline
$0$ & 1.80 & 10.002 & 5.31 & 0.46 & 3.01 \\ \hline
$1.0\times10^{-52}$ & 1.80 & 10.00 & 5.31 & 0.46 & 3.01 \\ \hline
$1.0\times10^{-50}$ & 1.80 & 10.00 & 5.31 & 0.46 & 3.01 \\ \hline
$1.0\times10^{-15}$ & 1.80 & 10.00 & 5.31 & 0.46 & 3.01 \\ \hline
$1.0\times10^{-14}$ & 1.80 & 10.00 & 5.30 & 0.46 & 3.01 \\ \hline
$1.0\times10^{-13}$ & 1.76 & 9.82 & 5.15 & 0.51 & 2.94 \\ \hline
$2.0\times10^{-13}$ & 1.73 & 9.66 & 5.02 & 0.56 & 2.87 \\ \hline
$5.0\times10^{-13}$ & 1.64 & 9.40 & 4.67 & 0.72 & 2.64 \\ \hline
$7.0\times10^{-13}$ & 1.59 & 9.25 & 4.48 & 0.85 & 2.47 \\ \hline
$7.5\times10^{-13}$ & 1.58 & 9.17 & 4.44 & 0.88 & 2.42 \\ \hline
$8.0\times10^{-13}$ & 1.57 & 9.14 & 4.40 & 0.92 & 2.38 \\ \hline
$8.5\times10^{-13}$ & 1.56 & 9.06 & 4.37 & 0.96 & 2.34 \\ \hline
$9.0\times10^{-13}$ & 1.55 & 8.98 & 4.33 & 1.00 & 2.29 \\ \hline
\end{tabular}%
\end{center}
\end{table}
%%%%%%%%%%%%%%%%%%%%%%%%%%%%%%%%%%%%%%%%%%%%%%%%%%%%%%%%%
%
In this table (Table \ref{tab1}) it is seen that for $\Lambda =10^{-52}m^{-2}
$ (from the cosmological perspective, the amount of cosmological constant is
about $10^{-52}m^{-2}$), there is no effect on the structure of hybrid
neutron star. In addition, for the cosmological constant in the range $%
10^{-14}m^{-2}\geq \Lambda \geq 10^{-52}m^{-2}$, this parameter does not
have significant effects on the structure of hybrid neutron stars. But for $%
\Lambda >10^{-14}$ $m^{-2}$, the maximum mass and radius of hybrid neutron
star decrease when $\Lambda $ increases. Here, we can ask this question:\ "%
\textit{why do the maximum mass of hybrid neutron stars decrease when the
value of cosmological constant increases?".} We will answer this question
after evaluating the effects of $\Lambda $ on the other properties of hybrid
neutron stars.

Now, we want to study the other properties of hybrid neutron stars such as
Schwarzschild radius, average density, gravitational redshift, and
Buchdahl-Bondi bound in the presence of $\Lambda $. We calculate these
properties as follows.

The modified Schwarzschild radius in Einstein-$\Lambda $ gravity is obtained
as \cite{BordbarHE} 
\begin{equation}
R_{Sch}=\frac{[(3GM+\sqrt{\frac{c^{4}}{\Lambda }+9G^{2}M^{2}})\Lambda
^{2}c]^{\frac{1}{3}}}{\Lambda c}-\frac{c}{[(3GM+\sqrt{\frac{c^{4}}{\Lambda }%
+9G^{2}M^{2}})\Lambda ^{2}c]^{\frac{1}{3}}},  \label{eq20}
\end{equation}%
for $\Lambda =0$, the above equation reduces to the Schwarzschild radius in
Einstein gravity \cite{Schwarzschild}. $M$ is the mass of a hybrid neutron
star. The modified Schwarzschild radius depends on the maximum mass and
cosmological constant (see Table \ref{tab1}). We see that the modified
Schwarzschild radius decreases by increasing the cosmological constant.

Another quantity that we have brought in Table \ref{tab1} is the
gravitational redshift in the presence of the cosmological constant from the
following formula \cite{BordbarHE}, 
\begin{equation}
z=\frac{1}{\sqrt{1-\frac{2GM}{c^{2}R}-\frac{\Lambda }{3}R^{2}}}-1.
\label{eq22}
\end{equation}

Our calculations are presented in Table \ref{tab1}. We find that the
gravitational redshift increases with increasing $\Lambda $. According to
the fact that the gravitational redshift cannot be more than $1$ for compact
objects, we find an upper limit on $\Lambda $. Our results indicate that
there is the upper limit on the cosmological constant for $\Lambda <9\times
10^{-13}m^{-2}$.

The Buchdahl-Bondi bound is another quantity that we have to respect. The
modified Buchdahl-Bondi bound in Einstein-$\Lambda $ gravity is given by $%
M\leq M_{BB}$ \cite{Mak}, where 
\begin{equation}
M_{BB}=\frac{2c^{2}}{9G}R-\frac{\Lambda c^{2}}{3G}R^{3}+\frac{2c^{2}}{9G}R%
\sqrt{1+3\Lambda R^{2}}.  \label{eq22mb}
\end{equation}

This relation reduces to $M\leq \frac{4c^{2}}{9G}R$ for $\Lambda =0$ \cite%
{Buchdahl1,Bondi,Buchdahl2}. The calculated results of $M_{BB}$ are
presented in Table \ref{tab1}. These results indicate that the obtained
stars respect to this bound.

\section{Dynamical stability}

\begin{figure*}[tbh]
\centering
\includegraphics[width=0.6\linewidth]{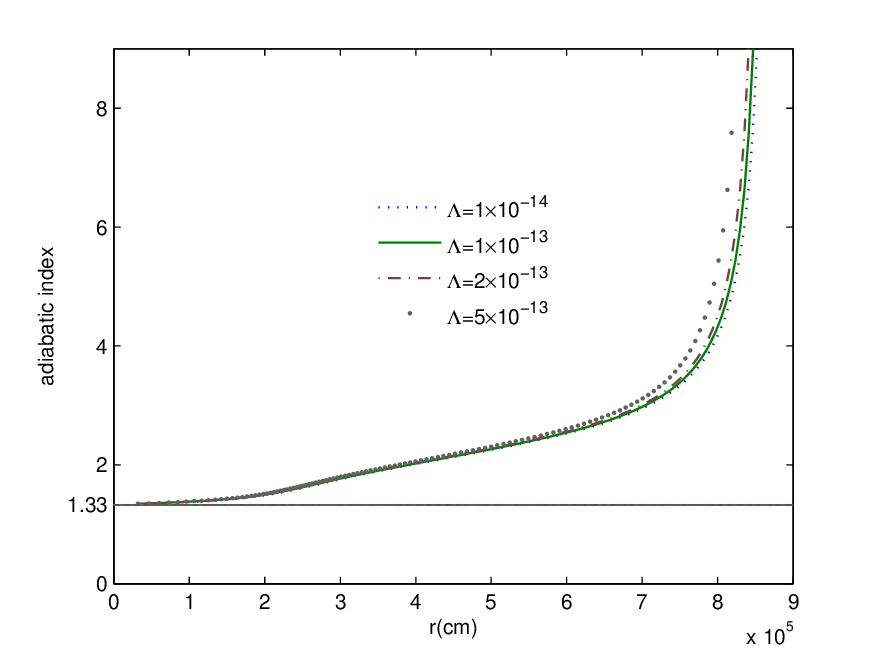}
\caption{Adiabatic index versus radius for different $\Lambda$.}
\label{gama}
\end{figure*}

To consider dynamically stability condition the adiabatic index $(\gamma )$
is plotted in radius parameter in Fig. \ref{gama}. This condition (which is $%
\gamma >\frac{4}{3}$) implies that stars are stable against the radial
adiabatic infinitesimal perturbations \cite%
{Chandrasekhar,Bardeen,Knutsen,Kalam}. The $\gamma $ can be obtained by the
following relation, 
\begin{equation}
\mathbf{\gamma =}\frac{\rho c^{2}+P}{c^{2}P}\frac{dP}{d\rho }.  \label{eq23}
\end{equation}
%
%%%%%%%%%%%%%%%%%%%%%%%%%%%%%%%%%%%%%%%%%%%%%%%%%%%%%%%%%%%%%%%
%%%%%%%%%%%%%%%%%%%%%%%%%%%%%%%%%%%%%%%%%%%%%%%%%%%%%%%%%%%%%%%

In addition to the above approaches, we investigate the dynamical stability
by using Shapiro and Teukolsky's point of view \cite{Shapiro}. They believe
that for dynamical stability, the pressure-averaged value of adiabatic index 
$\overline{\gamma }=\frac{\int_{0}^{R}\gamma pr^{2}dr}{\int_{0}^{R}pr^{2}dr}$
, must be greater than $\frac{4}{3}$. The determined results of the average
value of $\overline{\gamma }$ for different cosmological constants are as
follows 
\begin{eqnarray}
\overline{\gamma }_{{\Lambda }=7\times 10^{-13}} &\mathbf{=}&2.480,  \notag
\\
&&  \notag \\
\overline{\gamma }_{{\Lambda }=5\times 10^{-13}} &=&2.700,  \notag \\
&&  \notag \\
\overline{\gamma }_{{\Lambda }=2\times 10^{-13}} &=&2.695,  \notag \\
&&  \notag \\
\overline{\gamma }_{{\Lambda }=1\times 10^{-13}} &=&2.694,  \notag \\
&&  \notag \\
\overline{\gamma }_{{\Lambda }=1\times 10^{-14}} &=&2.686.
\end{eqnarray}

The obtained results in Fig. \ref{gama}, and value of $\overline{\gamma }$,
show that by applying the effect of the cosmological constant to hybrid
neutron stars, these objects are stable against the radial adiabatic
infinitesimal perturbations.

\section{Other properties: Contraction}

According to our results in the Table \ref{tab1}, and Fig. \ref{radius}, we
find that, when the structure of the hybrid neutron stars are calculated
with HEE in Einstein-$\Lambda $ gravity theory, the radius of these stars
reduces with increasing the cosmological constant, $\Lambda $, and therefore
the stars are contracted. To see better contraction of these stars due to
the effect of cosmological constant, we have brought the result of the
radius of the hybrid neutron stars with $M=1.4M_{\,\odot }$ in Table \ref%
{tab2}. Just as this table shows, the radius of hybrid stars with
gravitational mass equals $1.4M_{\,\odot }$, decrease with increasing $%
\Lambda $. 
\begin{table*}[tbp]
\caption{Radius and other properties of the hybrid neutron stars with $%
M=1.4M_{\,\odot }$ for different $\Lambda $ .}
\label{tab2}
\begin{center}
\begin{tabular}{||c|c|c|c|c||}
\hline\hline
$\Lambda (m^{-2}) $ & $R\left(km\right) $ & $\ \ \ \sigma$ & $K(10^{-8}$ $%
m^{-2})$ & $z$ \\ \hline\hline
$0$ & 10.76 & 0.38 & 1.15 & 0.270 \\ \hline
$1.00\times10^{-52}$ & 10.76 & 0.38 & 1.15 & 0.274 \\ \hline
$1.00\times10^{-14}$ & 10.76 & 0.38 & 1.16 & 0.278 \\ \hline
$1.00\times10^{-13}$ & 10.42 & 0.39 & 1.43 & 0.327 \\ \hline
$1.46\times10^{-13}$ & 10.34 & 0.39 & 1.53 & 0.350 \\ \hline
$2.00\times10^{-13}$ & 10.20 & 0.40 & 1.67 & 0.379 \\ \hline
$5.00\times10^{-13}$ & 9.76 & 0.41 & 2.36 & 0.547 \\ \hline
\end{tabular}%
\end{center}
\end{table*}
%%%%%%%%%%%%%%%%%%%%%%%%%%%%%%%%%%%%%%%%%%%%%%%%%%%%%%%%%
%
For more details see Fig. \ref{radius}. From this figure, we see that the
hybrid neutron stars with $M=1.4M_{\,\odot }$ have different radii. In other
words, by increasing the value of the cosmological constant, the radius of
these stars decreases. At the same time, their mass is constant ($%
M=1.4M_{\,\odot }$), which confirms the existence of a contraction.

\subsection{Strength of Gravity}

To evaluate the strength of gravity, we investigate the compactness and the
Kretschmann scalar of these compact objects in the presence of the
cosmological constant in the following parts.

\textbf{Compactness}: For a spherical object, the compactness of the star
may be defined as $\sigma =\frac{R_{Sch}}{R}$. Considering this definition
of compactness, we calculate it for different values of the cosmological
constant in Table \ref{tab2}. Our results show that the strength of gravity
increases when the value of the cosmological constant increases.

\textbf{Kretschmann scalar}: Notably, we can study the Kretschmann scalar to
measure curvature in a vacuum. 
\begin{equation}
K=\sqrt{R_{\mu \nu \gamma \delta }R^{\mu \nu \gamma \delta }}=2\Lambda \sqrt{%
\frac{2}{3}}+\frac{4\sqrt{3}GM}{c^{2}R^{3}}.
\end{equation}

As we can see in Table \ref{tab2}, the Kretschmann scalar increases by
increasing the cosmological constant.

Here our analysis of compactness, Kretschmann scalar and redshift of hybrid
neutron stars in the presence of the cosmological constant show that by
increasing $\Lambda $, these quantities increase, whereas the radius of
these stars decreases. In other words, these stars become very compact when $%
\Lambda $ increases.

According to the above results, now we can answer the above question "%
\textit{why do the maximum mass of hybrid neutron stars decrease when the
value of the cosmological constant increases?}" \textit{Answer:} It is known
that from the cosmological point of view, by adding the cosmological
constant to Einstein's gravity, this theory can explain the accelerated
expansion of our Universe. In other words, this constant creates a pressure
or repulsive force inside of universe (the large scale), which leads to this
acceleration. According to our results, by increasing $\Lambda $, the hybrid
neutron stars became very compact. Indeed, the cosmological constant inside
of hybrid neutron stars or compact objects (the small scale) may act as an
attractive force (opposite of effect it on universe). As we know, for having
a balance between attractive force (gravitational force due to mass of
compact object $+$ the cosmological constant) and repulsive force (internal
pressure), the maximum mass of compact objects decreases by increasing the
strength of gravity (or by increasing the value of the cosmological
constant).

\section{Summary and Conclusion}

In this work, we checked the effect of the cosmological constant on the
structure of neutron stars with quark matter (hybrid neutron stars). Because
of high densities in the interior of neutron stars, we believe that these
compact stars contain a deconfined quark phase. Therefore, we consider a
crust of hadronic matter, a mixed phase of quark and hadronic matters, and a
quark core for a neutron star. To calculate the structure of the hybrid
neutron stars, we used the HEE in the Einstein-$\Lambda $ gravity that is a
modified TOV equation. By numerically integrating this HEE, we determined
these stars' gravitational mass and radius. The results showed that for $%
\Lambda $ smaller than $10^{-14}$ $m^{-2}$ (especially $\Lambda =10^{-52}$ $%
m^{-2}$, the cosmological constant from the cosmological perspective), $%
\Lambda $ does not affect the structure of hybrid neutron star. But for $%
\Lambda >10^{-14}$ $m^{-2}$, the structure of this star is affected by the
cosmological constant. Our results showed that hybrid neutron stars' maximum
mass and radius decrease by increasing the cosmological constant, $\Lambda $%
. The compactness, the Kretschmann scalar, and gravitational redshift
increase with $\Lambda $. Therefore we can conclude that a hybrid neutron
star is contracted when the cosmological constant increases.

Another exciting result was related to dynamical stability. By applying the
effect of the cosmological constant to these stars, the dynamical stability
is kept inside them. In addition, we found an upper limit on the
cosmological constant which was for $\Lambda <9\times 10^{-13}m^{-2}$.

%%%%%%%%%%%%%%%%%%%%%%%%%%%%%%%%%%%%%%%%%%%%%%%%%%%%%%%%%%%%%%%%%%%%%%%%%%%%%%%%%%%%%%%%%%%%%%%%%%

\section*{Acknowledgements}

TY wishes to thank the Research Council of Islamic Azad University, Bafgh
Branch. GHB thanks Shiraz University Research Council. BEP thanks University
of Mazandaran. The work of BEP has been supported by University of
Mazandaran by title "Evolution of the masses of celestial compact objects in
various gravity".

%%%%%%%%%%%%%%%%%%%%%%%%%%%%%%%%%%%%%%%%%%%%%%%%%%%%%%%%%%%%%%%%%%%

%%%%%%%%%%%%%%%%%%%%%%%%%%%%%%%%%%%%%%%%%%%%%%%%%%%%%%%%%%%%%%%%%%%

%%%%%%%%%%%%%%%%%%%%%%%%%%%%%%%%%%%%%%%%%%%%%%%%%%%%%%%%%%%%%%%%%%%%%%%%%%%%%%%%%%%%%%%%%%%%%%%%%%%%%%

%%%%%%%%%%%%%%%%%%%%%%%%%%%%%%%%%%%%%%%%%%%%%%%%%%%%%%%%%%%%%%%%%%%%%%%%%

\end{document}